# Blockchain Applicability for the Internet of Things: Performance and Scalability Challenges and Solutions

**Ziaur Rahman** [1], **Xun Yi** [1], **Sk. Tanzir Mehedi** [2], **Rafiqul Islam** [3], **and Andrei Kelarev** [1]

[1]  School Science, RMIT University, Melbourne 3001, Australia; xun.yi@rmit.edu.au (X.Y.); andrei.kelarev@rmit.edu.au (A.K.)
[2]  Department of Information and Communication Technology, Mawlana Bhashani Science and Technology University, Tangail 1902, Bangladesh; tanzirmehedi@ieee.org
[3]  School of Computing, Mathematics and Engineering, Wagga Wagga, NSW 2678, Australia; mislam@csu.edu.au
*   Correspondence:  s3677291@student.rmit.edu.au;  Tel.:  +61-0426-117-006

**Abstract:** Blockchain has recently drawn wide attention in the research community. Since its emergence, the world has seen the expansion of this new technology, which was initially developed as a digital currency more than a decade ago. A self-administering ledger that ensures extensive data immutability over a peer-to-peer network has made it attractive for cybersecurity applications, including sensor-enabled Internet of Things (IoT) systems. Brand new challenges and questions now demand solutions, as IoT devices are now online in a distributed fashion to simplify our everyday lives. Motivated by those challenges, the work here has detailed issues from which an IoT infrastructure can suffer if the wrong blockchain technology is chosen. Unlike a typical review, this paper focuses on security challenges of the blockchain-IoT eco-system through critical findings and applicable use cases. The contribution directs Blockchain architects, designers, and researchers in the domain to select an unblemished combination of Blockchain-powered IoT applications. In addition, the paper provides insight into the state-of-the-art Blockchain platforms, namely Ethereum, Hyperledger, and IOTA, to exhibit their respective challenges, their constraints, and their prospects in terms of performance and scalability.

**Keywords:** blockchain; Hyperledger; Ethereum; distributed ledger; Internet of Things; public consensus; scalability

## 1. Introduction

The integration of blockchain (BC) with IoT has shown immense effectiveness and potential for future improvements in scalability and productivity. Therefore, how these emerging technologies can be deployed together to secure end-to-end and sensor-embedded automated solutions while ensuring their scalability and productivity has become a key priority. The world has benefited from the adaptations of different heterogeneous IoT solutions, ranging from healthcare to transportation systems [61]. The existing centralized edge and fog-based IoT infrastructure/applications may not be secure, scalable, and efficient enough to address larger enterprise challenges. Most existing IoT solutions are concerned with the network of sensor-enabled smart appliances, which permits physical device services on the cloud [61]. An immutable timestamp ledger is used for distributed data, including payment, contract, personal data storing, data sharing, and healthcare systems, due to its salient features such as immutability, a distributed structure, consensus-driven behavior, and transparency [62].

There are various reasons why BC technology may be highly promising for assuring the efficiency, scalability, and security of the heterogeneous IoT setup. The issues related to the emerging IoT networks and several BC roles can be solved as follows. Firstly, approximately 50 billion devices will be connected by 2022 [63]. Several efforts have been made to reveal the associated challenges [64]. In response to the adaptability of trillions of devices in the near future, it should not be difficult to handle using decentralized BC technology. As BC requires no centralized database and addresses are directly addressable, one device can directly send information to another [65]. This means that this technology has limitless and scalable registration capability. The second issue is how to control a large number of devices on a distributed and decentralized platform. In response, BC technology provides open peer-to-peer connectivity for intra-device communication between physical or virtual appliances [63]. The third one is how it provides compliance and legitimate governance for all autonomous systems involved. In response, BC technology has a smart contract-based immutable open ledger system. Therefore, transparency is a characteristic of this technology that ensures more comprehensive autonomy and trustworthy governance [66]. The last concern is how BC technology can address the security complexities of the new heterogeneous IoT ecosystem that is rapidly emerging and evolving. Since 2008, Bitcoin has been adapting to ongoing Internet challenges [62]. Apart from financial transactions, it has shown immense potential in the field of IoT, incorporating features such as an elliptic curve digital signature algorithm (ECDSA) [67], the zero-knowledge proof (ZKP) [68], message signing, differential privacy [65], cryptographic message verification, and others.

The goal of this research is to identify the trade-offs that the heterogeneous IoT ecosystem typically faces due to the wrong choice of BC technology. Unlike a survey or review, the findings of this research are aimed at solving particular performance and scalability issues in a BC-enabled IoT architecture. The contribution directs developers and academics in this field to select the best BC-enabled IoT applications. The claimed contributions are justified through the respective sections of the paper. We discuss the suitability of BC to eliminate the problems that emerge from BC and IoT integration [69]. We also explain how existing solutions, namely, the Microsoft (MS)-Azure IoT workbench and the IBM IoT architecture, adopt different BC platforms such as Bitcoin (BTC), Ethereum (ETH), Hyperledger (HLF), and Kovan. The following section illustrates BC's potential for specific IoT issues. The challenges come to light through a use case analysis, where a sensor-enabled system finds appropriate devices, manages access control, and supports the compliance of smart contracts. In addition, this research supports the use of smart contracts in IoT systems and points out possible flaws in data integrity, scalability, and confidentiality.

The research paper is organized as follows: Section 2 discusses the related work in this field. Section 3 discusses the internal design of BC technology and the specialized categories within which it can be applied. The suitability of BC technology for IoT applications with comparative analysis and contemporary technologies, including HLF, IOTA, and the MS-Azure IoT architecture is discussed in Section 4. Section 5 summarizes with a brief table and graphs showing the challenges and proposed solutions at a glance as well as their applicability concerning throughput, latency, and execution time. Section 6 discusses a set of use cases where BC technology is an inevitable peer of the IoT discussed herein. Finally, Section 7 and 8 include a discussion and summary and provide feasible future directions and theoretical and practical implications, respectively.

## 2. Related Work

Apart from the financial domain, BC technology has been showing its far-reaching prospects in different application areas since its first emergence in 2008 [70]. Since what is written cannot be modified, the nature of the BC ledger, in addition to the pseudo-anonymous, traceable peers over the transparent distributed network, makes BC an indispensible tool in IoT [71]. The field includes smart areas, grids, vehicles, industry, the supply chain, food and drug safety, the e-commerce of agricultural products, medical technology, and industrial predictive maintenance [72]. Significant research activity has also been done in digital data copyright protection of, ID verification, real state land ownership transfer, smart-taxation immigration, electronic voting, and privacy-principle compliance [73]. Even in an IT sector such as Blockstack [74], BigchainDB [75] utilizes the BC smart-contract and consensus mechanism.

Namecoin incorporates a distributed hash table (DHT) that communicates with the virtual chain after separating BC dApps, operations, and off-chain storage entities [76]. It hashes the name data tuples, state transitions, and records in an on-chain BC ledger, and the DHT stores the payload, digital data, and associated signatures. However, some researchers have employed the immense benefits of IPFS for storing access control and compliance data [77]. They proposed customizing attribute-based encryption after replacing the centralized cloud-dependency by leveraging the public chain, namely Ethereum. In line with that, BigchainDB employs a Tendermint-distributed database based on the weak synchronization of the BC engine deployed on the Byzantine consensus (BFT) [78]. This promising data and execution project enables large-scale and real-time data protection and management for industrial IoT security and privacy.

The rapid growth of employing IoT sensors encounters several challenges, such as data protection, analytical management, and storing voluminous real-time data [79,80]. NoSQL, or the Hadoop repository, initially attracted researchers in the IoT domain but was not convincing because of its centralized structure, its single point of failure (SPOF) nature, and its security issues [81]. Based on this, the authors proposed an approach after attaching multiple cloud-centric database models that were promising [80]. However, various dependencies lead to SPOF, trust, and security intricacies. Several comprehensive works have proposed an Edge solution to address such challenges, which enormously motivated the idea we introduce here on BC technology. However, besides miners' high-energy incentive disputes, the Blockchain network encounters scalability issues that some works [82,83] have concentrated on and aimed at solving through plausible remedies [84]. Some of the demonstrations, including channel-driven communication between the data owner and requester using shared secret keys [85], as well as BC for trusted computing, utilized the underlying public Blockchain (i.e., BTC and ETH) to provide a miner's network that emulates a trusted server. However, apart from the potential threat of leaking secure, shared secrets, establishing a secure channel without consortium BC (i.e., HLF and Corda) seems untrustworthy.

In August 2018, focusing on security and privacy, a group of authors proposed applying multi-signatures and BC for decentralized energy trading [84,86]. Following the same motivation of multi-signature and consortium BC, the authors improved their P2P vehicle trading mechanism to the IIoT energy trading system in September 2018.

The certificate-less cryptography was initially introduced to abolish the key IBE escrow issues in the early years of this century; however, several works in the following years worked toward its efficient improvement [87,88]. From an IoT perspective, multi-signature based certificate-less authentication reduces computational costs and signing latency, especially for a network involving light-weight sensors [89]. Considering the key dissipation hardship, costs, and latency, one of the latest works portrays a convincing resolution upon aggregating the Edge and DHT. The works claimed to be suitable for industrial IoT but lack details on how it overcomes the public BC network deployment and delay in the transaction (TX) generation, verification, and broadcasting [90]. Moreover, the adaption and construction of a key distribution center (KDC) attempted to extenuate system performance toward a centralized architecture [91,92]. Table 1 concludes the overview of the selected recent literature reviews on BC and BC-based IoT applications.

**Table 1.** Overview of the selected recent literature reviews on BC and BC-based IoT applications.

| Ref. | Year | Research Area | Summary Contributions and Features |
|------|------|---------------|-------------------------------------|
| [71] | 2017 | BC for CPS | Resilience of Interacting distributed energy at speed, scale and security with blockchain |
| [93] | 2017 | BC Improvement | Scaling PBFT agreements for further improvment of Bitcoin |
| [94] | 2017 | IoT Security | SecKit: a model-based security toolkit for the internet of things |
| [95] | 2018 | BC-based IoT security | A Review, blockchain solutions, and open challenges |
| [88] | 2018 | BC for Cloud Security | How to adapt BC for securing Cloud |
| [96] | 2018 | Public BC for Security | A Special Model called RapidChain for fast Protocol using full Sharding methods |
| [80] | 2018 | BC for Iot Security | How BC could be applied for a large scale IoT System focusing data storage and protection. |
| [97] | 2019 | BC Consensus on PBFT | How Practical Byzantine is more efficient that PoW or PoS |
| [98] | 2019 | Permissioned BC | Showing the immense prospects of Hyperledger Fabric for distributed system |
| [99] | 2020 | BC Access Control for IoT | BC has verified features for scalable access management of IoT |
| [78] | 2020 | BC for Data Management | BC based data maintenance with identity management |
| [100] | 2021 | BC for Access Control | An extended model for Access control using Permissioned Blockchain |
| [101] | 2021 | BC for Access Control | Data Accountability and Provenance Tracking using BC |
| [86] | 2021 | BC-based Security Framework for CPS | Blockchain-Based Security Framework for a Critical Industry 4.0 Cyber-Physical System |
| [72] | 2022 | BC-based AI-enabled CPS | Blockchain based AI-enabled Industry 4.0 CPS Protection against Advanced Persistent Threat |

## 3. Preliminaries of BC Technology

The BC's main task is to replace traditional and trust-created intermediaries with distributed systems to solve common trust issues [65]. It also helps in forming a permanent and transparent record of the exchange of processing and avoids the need for an intermediary. Instant value exchange, decentralized value exchange, and pseudonymous value transfer are all terms used to describe BC technology [102]. It also makes sure that the ledger building preserves a set of transactions shared among all participating nodes, which needs to be necessarily verified and validated by others [65,103]. Joining brand new transactions is commonly referred to as "mining" and requires the solution of a sophisticated and large computational problem, which in nature is complex but is easy to authenticate using a selected consensus algorithm in a network of untrusted and anonymous nodes. The consensus algorithm requires a significant amount of resources in order to ensure that only authorized blocks may join the network. In addition, the communication between nodes is encrypted using changeable public keys (PKs) to avoid monitoring, which has attracted attention in non-monetary applications [66]. Moreover, the hash of the previous block, the timestamp, the transaction root, and the nonce generated by the miner are also seen in a sample chain of blocks, which makes the BC more secure [104]. Figure 1 shows an overview of different blocks with timestamp, hash, nonce, and transaction data. Therefore, using BC-enabled applications has become much more transparent because of this development.

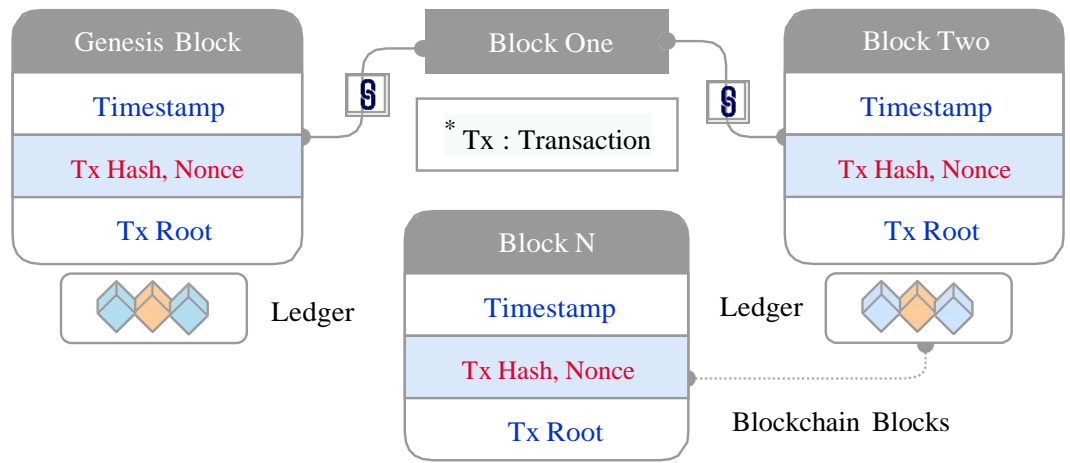

**Figure 1.** Overview of different blocks with timestamp, hash, nonce, and transaction data.

Using high-security smart devices and smart technologies for authentication to ensure seamless communication, decentralized data processing, or even autonomous systems for data purchase and others may demonstrate its promise in this field [105]. Consequently, the IoT devices might be equipped with the Internet to make every part of human life more convenient and less tedious [106,107].

### 3.1. Category of BC Technology

In this section, we have covered three different approaches to BC technology: public ledger-based, private ledger-based, and protected ledger-based. A comparative categorization of BC ledgers is shown in Figure 2 based on the accessibility of the considered ledger.

### 3.1.1. Public Ledger-Based BC

In public ledger-based BC technology, anyone can transmit, verify, and read transactions on the network, as well as obtain and run the scripts necessary to participate in the BC mining process using several consensus methods, making it known as a "permission-less" BC technology [102]. Even though any anonymous user may transmit, view, and authenticate an incognito transaction, it offers the highest level of anonymity and transparency [108]. ETH [109] and BTC are two of the most common examples of public BC technology.

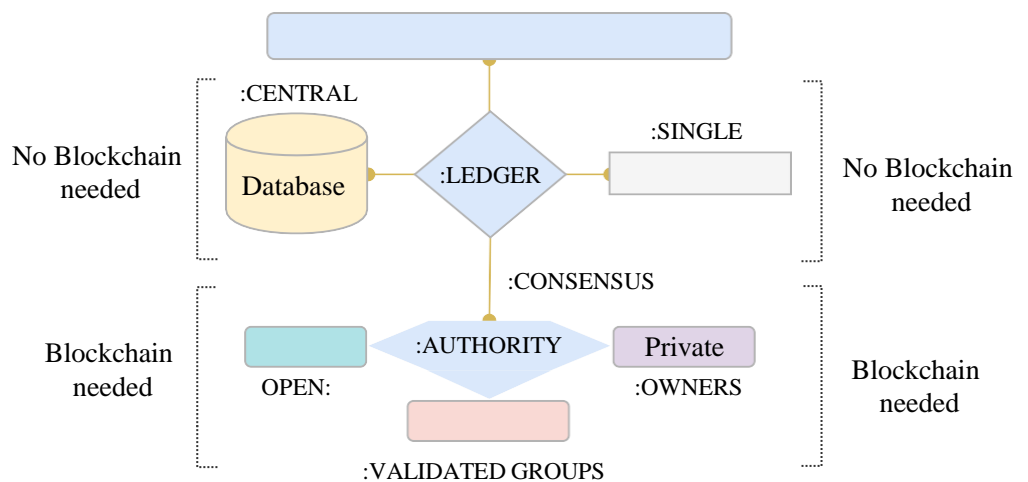

**Figure 2.** The classification of BCs according to the requirement analysis.

### 3.1.2. Private Ledger-Based BC

Private ledger-based BC does not require a consensus mechanism or mining to provide anonymity because it restricts read and modification rights to a certain organization. The read authority is sometimes restricted to an arbitrary level, but most of the time, transaction editing is rigorously permissioned [110]. Private-typed BC approaches might be stated to be used in the ledger-building process for coins controlled by Eris and Monax or the Multichain [103]. To cite an example, a permissioned-based BC technology like Quorum is now available on ETH, though ETH itself is a public ledger-based BC technology.

### 3.1.3. Protected Ledger-Based BC

The protected ledger-based BC is also known as a consortium/federated, hybrid, or public-permissioned BC, which is run and maintained by a group of owners or users [108]. Protected ledger-based BCs include HLF by the Linux Foundation [61] and IBM and R3 with Corda or the Energy Web Foundation [110]. Moreover, if the authority is restricted within a validated group, then protected ledger-based BC seems to be more suitable than a public or private ledger-based BC system [111].

Moreover, Figure 2 shows that, if the system has a centralized or single ledger system, no category of the BC is needed there. Additionally, we discussed the performance comparison between IOTA and the other BC technologies. According to the IOTA team, its ledger is a public permission-less backbone for IoT applications [107]. This means it will enable transactions between connected devices, and anyone on the network can access its ledger.

## 4. Suitability of BC Technology for IoT Applications

Although BC technology is capable of solving all IoT-related issues, there are a few situations where a centralized database is preferable. BC-based use cases need to be explored before being implemented in this area.

### 4.1. Comparison of Several Consensus Protocols

Table 2 describes the comparison among different popularly used consensus mechanisms for BC technology. It shows that Proof-of-Work (PoW) and Proof-of-Stake (PoS) need more computational resources in contrast to Byzantine fault tolerance (BFT) and proof-of-authority (PoA), which outperform their peers. However, BFT and PoA are both difficult to adjust [112]. Even though they have dependencies, they seem to work for IoT nodes. For scalability and overhead, blocks needed to be verified by all nodes available in the network, with a quadratic increase in traffic and a disobedient overhead of data processing power, which requires many expandable IoT devices (e.g., LORA) with limited bandwidth [105]. IoT devices tend to fail with higher delays, but BTC takes nearly 30 min to finalize a transaction. It also has security overhead, making it inapplicable for IoT [113]. Because of the interaction between IoT nodes, the throughput of BTC (7/transaction) will push it over the limit. As a result, many people have switched from BTC to BFT-based HLF or non-consensus-driven systems such as IOTA [61,102]. The applicability of different BC-based systems depends on whether consensus and non-consensus approaches are discussed.

**Table 2.** A comparison of several widely used consensus techniques for BC technology.

| Attributes | PoW | PoS | BFT | PoA |
|---|---|---|---|---|
| Category | Public | Pub/Protected | Private | Protected |
| Throughput | Little | Big | Big | Big |
| Random | No | Yes | No | No |
| P-Cost | Has | Has | Not | Not |
| Token | Has | Has | Not | Native |
| Trust | Trust-less | Trust-less | Semi | Trusted |
| Scalability | Big | Big | Little | Medium |
| Reward | Yes | No | No | No |
| Example | BTC | ETH | HLF | Kovan |

### 4.2. Comparative Analysis of ETH, HLF, and IOTA Technology

ETH technology, launched with the intention of competing with BTC, is a flexible BC platform with a required smart-contract and PoW consensus mechanism known as *Ethash*, which generates the probabilistic hash using directed acyclic graphs (DAGs) [66]. It greatly helps with extensive IoT applications and some of its efficiency trade-offs. ETH needs almost 20 s to open a new block after mining, as *Ethash* works based on the PoW mechanism [102].

HLF is an authenticated and encrypted type of BC technology. It applies authentication widely, along with chain-code-based smart contracts and consensus with existing practical Byzantine fault tolerance (PBFT) [67]. Anchors of trust are added to the asymmetric cryptographic technique and the digital signature qualities with SHA3 or ECDSA as an additional feature of the system [102]. A self-execution capacity such as asset or resource transfer across network peers is required for its implementation of smart contracts. It has low latency with respect to other comparative distributed ledger implementations. Furthermore, according to IBM's Bluemix-Watson IoT design, which is shown in the next section, Fabric was selected as the BC medium.

IOTA is an unique distributed ledger that does not use an explicit BC at all; rather, it implements a DAG of transactions. In place of multi-block transactions, individual transactions are approved and two other transactions are implied [102]. IOTA tangles have the potential to be effectively integrated with the IoT in order to provide security and privacy.

Figure 3 shows a comparative analysis of ETH, HLF, and IoTA technology in terms of performance and scalability.

| BC Type | Consensus | Delay & SC | | |
|---|---|---|---|---|
| HLF | PBFT | 10-1000ms | Yes | High-computation inesive |
| ETH | Etash | 10k ms | Yes | Light computation-intensive, High network use |
| **IOTA** | No [DAG] | 10 ms | No | Light computation-intensive, Low network use |

**Figure 3.** The comparative analysis of ETH, HLF, and IOTA technology.

### 4.3. MS-Azure IoT Workbench

Figure 4 shows the Azure IoT framework, which, depending on the smart contract, streamlines client-side based applications for both web and mobile. It is used to validate, retrieve, and test programs or to consider novel use cases. A user interface is introduced for the end users to interact with it in different ways. Authenticated users can interact with the admin console, allowing them to use many functionalities, e.g., upload and deploy smart contacts depending on appropriate roles.

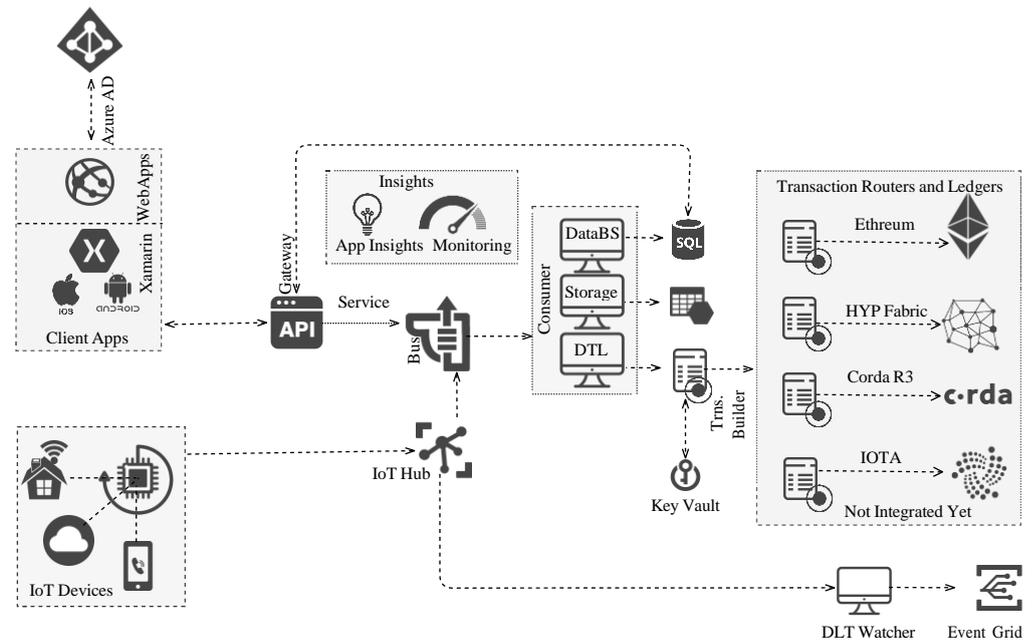

**Figure 4.** Azure IoT reference architecture that has been integrated with BC for securing IoT devices.

Figure 4 illustrates the REST API-based gateway service API used to replicate and send messages to an event broker, where the expansion of data into the BC technology is attempted. When data are requested, quarries are submitted to an off-chain database. Replicas of all chained meta-data and bulk-data that issue relevant configurations for smart-contract support are contained in the SQL database. Thus, developers can directly access the gateway servicing API to develop BC technology. Direct data submission to the service bus is an option for users who want their messages to be sent widely throughout the Azure infrastructure. As an example, this API may be used to build sensor-based tools or federated systems. In addition, there are several events hosted over the life of the application [102]. The gateway API or even the ledger's alerting trigger downstream code can accomplish this dependence on previous events. There are two types of event consumers that the MS-Azure consortium may locate [61]. The first one, enabled by the events, remains on the BC to access the off-chain SQL database. As a final response, it collects meta-data from API events related to document upload and storage. Figure 4 shows how the MS Azure IoT workbench becomes familiar with different BC frameworks. The MS Azure architecture may also be used to support HLF Fabric, Corda R3, and IOTA. The IoT Hub is connected to the IoT sensors through a bus, and the Transaction Builder is connected to this bus. Finally, in order to create a scalable and secure IoT device, an existing IoT workbench may be integrated with MS Azure.

*4.4. IBM BC-Integrated IoT Architecture.*

The IBM BC architecture for IoT solutions has three principal tiers, each with different roles [61]. Figure 5 shows a high-level IoT architecture that includes HLF Fabric as a BC service, Watson as an IoT Platform, and Bluemix as a cloud environment [68]. It can be divided into several components, as shown in Figure 5, showing its three layers, a service execution method, and the challenges it confronts. It also shows how IBM Blumix works. When executed, data gathered by smart devices and intelligent sensors are introduced to Watson using the ISO standard Message-Queuing-Telemetry-Transport (MQTT) protocol. Depending on the settlement, certain BC proxies are used to send data from Watson to the chain code of the HLF Fabric and are executed in the cloud.

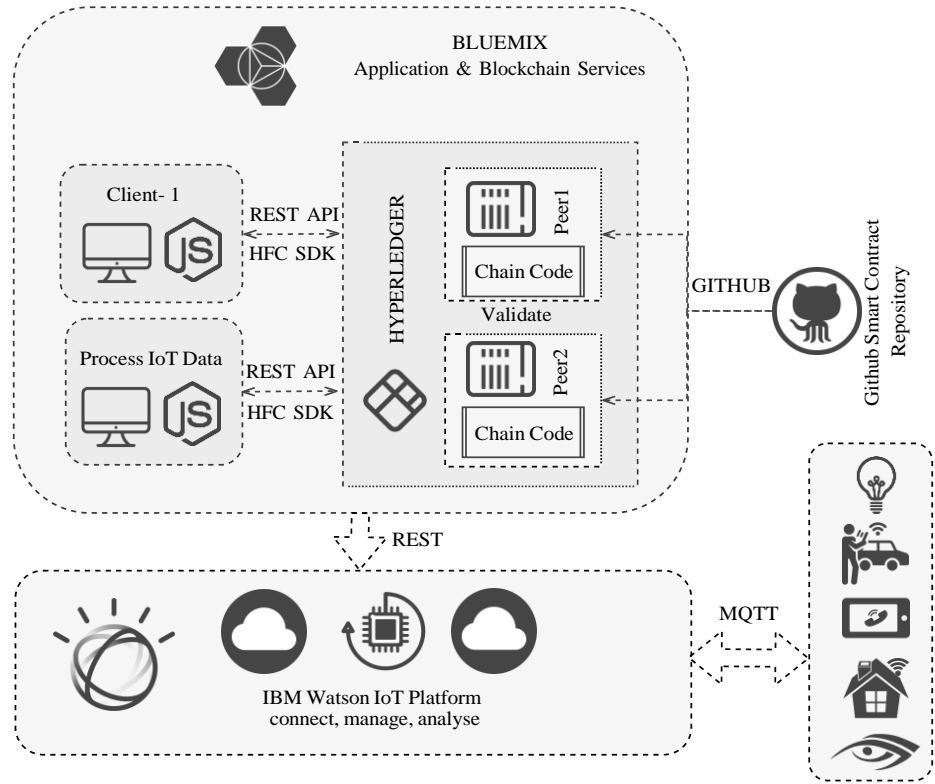

**Figure 5.** IBM Watson and Bluemix have been integrated with the IoT-BC service. Using Bluemix's BC network, Watson can communicate with IoT devices via Github's smart contact repository.

Furthermore, the HLF Fabric uses, instead of smart contracts, chain code written in Go. The desired business logic is elaborated by it and gives shape to the core distributed ledger solutions. Each transaction is preserved and prevailed, which is needed for BC transactions. Fabric contracts being chain-coded need certain APIs to run. As such, the chain code is in need of registration with services using any predefined APIs. Software development kits (SDKs) help developers to make Node.js applications that can maintain communication with BC networks. APIs are used to register and submit applications. The IBM BC-integrated IoT architecture on Bluemix provides many benefits to the distributed network, such as trust, autonomy, scalability, and security. There are many issues to be resolved. One of the important issues is hardware resources [66], since IoT devices are mostly low-powered devices and have less computation power. Therefore, encryption and transaction verification may use considerable electricity, which will increase both energy consumption and costs.

## 5. Challenges and Solutions for BC-Based IoT Applications.

Despite the many appealing features of BC for IoT applications, there are several challenges that must be addressed before successful adoption. The storage capacity, throughput, latency, execution time, privacy and security, and scalability of the BC-based IoT applications are addressed in this section. Following that, we have also thoroughly explained some inevitable challenges and their possible solutions. Figure 6 shows the challenges in BC-based IoT applications.

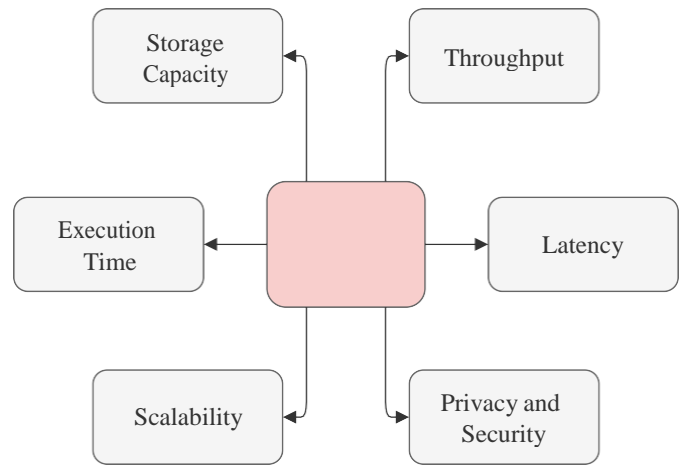

**Figure 6.** Challenges in BC-based IoT applications.

### 5.1. Challenges in Storage Capacity

As previously discussed, ETH and BTC have storage issues. Figure 7 shows how the storage capacity has been increasing day by day from 2015 to the first quarter of August 2021. The storage-intensive BC infrastructure is less suitable for heterogeneous IoT systems [114]. The massive amount of data generated by IoT devices raises the likelihood of a system crash due to the additional storage overhead [115]. In real-time heterogeneous IoT systems, ETH appears to be better suited for storage capacity than BTC, as shown in Figure 7. However, the storage capacity of a BC is not the only aspect that determines whether it is suited for heterogeneous IoT systems.

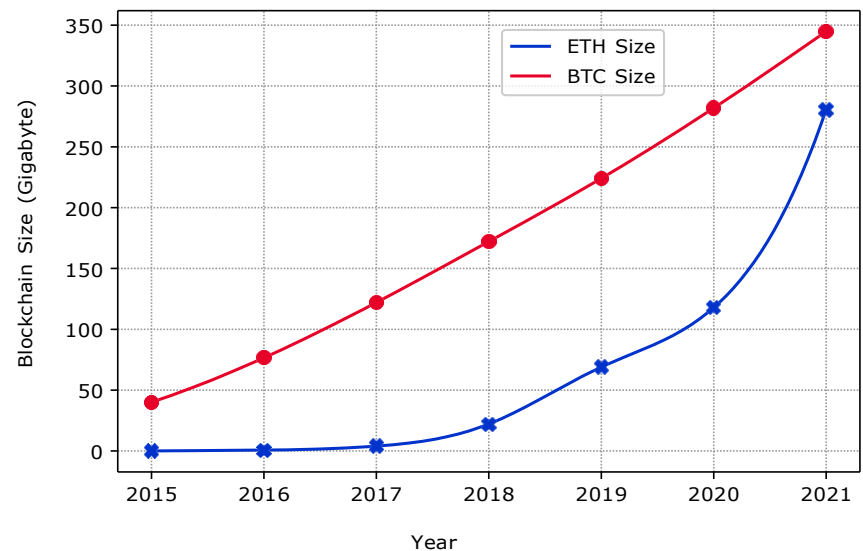

**Figure 7.** Storage capacity comparison between BTC and ETH technology using data from the Blockchain, Etherscan, and Statista websites.

### 5.2. Challenges in Throughput

Furthermore, we have considered the throughput of several BC technologies. Figure 8 compares the throughput of ETH, ETH Parity, and HLF Fabric in terms of the number of transactions per second, where HLF has the highest throughput for the Yahoo-Cloud-Serving-Benchmark (YCSB) and the Smallbank database. The datasets were found from [102], where they used the Blockbench framework to collect data. ETH Parity, on the other hand, has the lowest throughput compared to the others, implying that it is less appropriate for real-time heterogeneous BC-based IoT infrastructures.

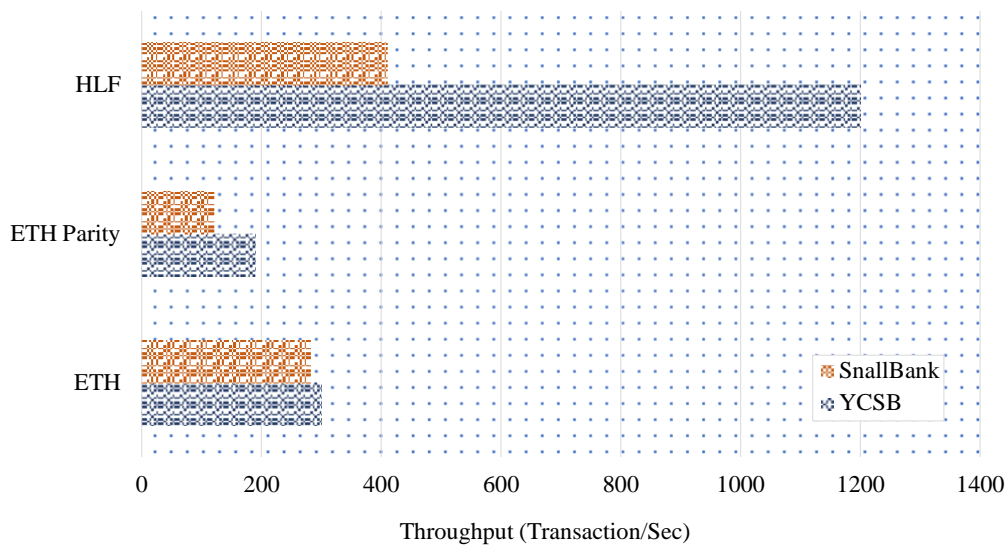

**Figure 8.** Throughput comparison between ETH, ETH Parity, and HLF Fabric.

*5.3. Challenges in Latency and Execution Time*

We have also considered the latency and execution time of several BC technologies. Figure 9 compares the latency and execution time of ETH, ETH Parity, and HLF Fabric, where HLF has the lowest latency and execution time for both databases. One of the ETH implementations, ETH Parity, is an alternative BC solution for IoT applications. Therefore, we considered both the ETH and ETH Parity to calculate latency and execution time. In addition, the Linux Foundation hosts the HLF, an open-source collaborative program aimed at improving cross-industry BC technology [107].

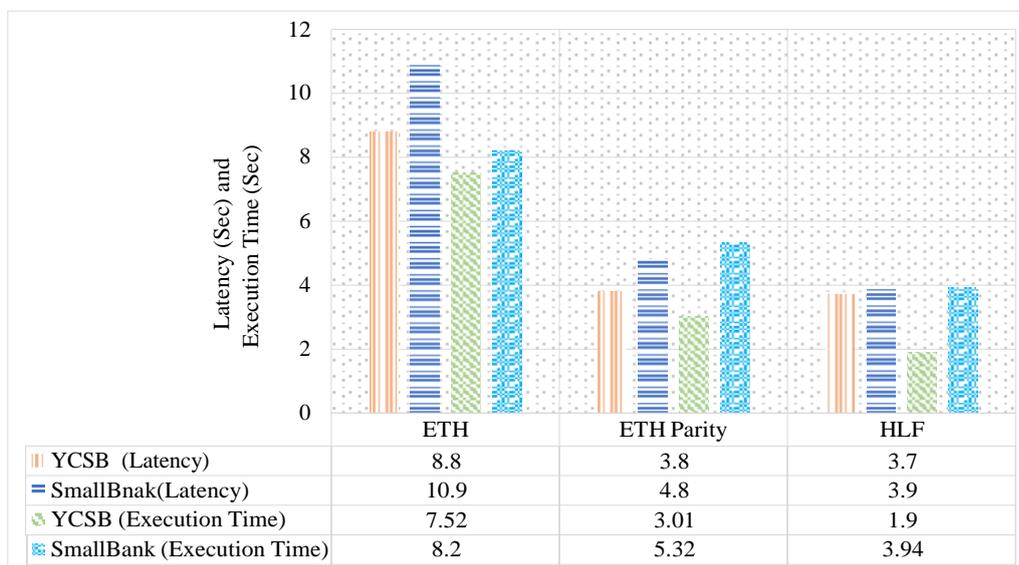

**Figure 9.** Latency and execution time comparison between ETH, ETH Parity, and HLF Fabric.

*5.4. Challenges in Privacy and Security*

BC technology works as a public ledger that secures and authenticates transactions and data through cryptography, which is more complex. With the rise and widespread adoption of BC technology, data breaches have become frequent. User information and data are often stored, mishandled, and misused, posing a threat to personal security and privacy. In terms of security, the data need to be tamper-proof, as some of the nodes may act maliciously or be compromised. As a result, proper security must be ensured before integration with the IoT infrastructure. Moreover, in terms of privacy, the data or transactions belong to various nodes in BC technology. Therefore, privacy needs to be ensured before integration with the IoT infrastructure.

*5.5. Challenges in Scalability*

Finally, we consider the scalability of several BC technologies. The scalability of BC technology is composed of node scalability and performance scalability. Node scalability in BC networks refers to the extent to which the network can add more participants without a loss in performance. Performance scalability refers to the number of transactions processed per second, impacted by the latency between transactions and each block size. A BC technology is considered scalable if it can add thousands of globally distributed nodes while still processing thousands of transactions per second. Currently, none of the existing BCs are appreciably scalable. Figure 10 shows a comparison of scalability, some of which are currently in use and some of which are in development.

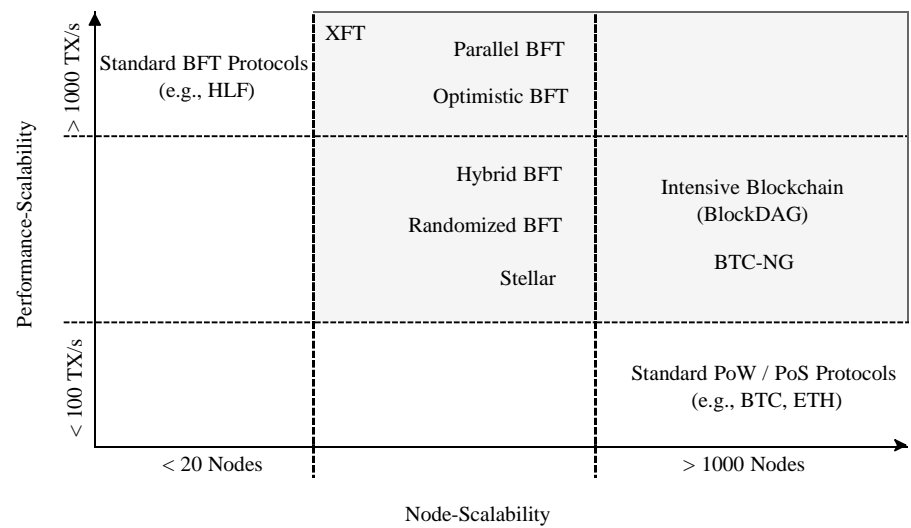

**Figure 10.** Scalability comparison between ETH, BTC, HLF, and other technologies still in development.

Public BCs such as BTC and ETH, by using PoW consensus mechanisms, have high node scalability and low performance scalability. On the contrary, HLF Fabric has low node scalability but high performance scalability. For heterogeneous IoT infrastructures of less than 20 nodes, this technology might be a viable solution. However, if we need more nodes, the amount of messaging that takes place between the nodes in PBFT can lower transaction throughput significantly. Therefore, a large-scale IoT system will be unable to successfully integrate with BC technology unless these challenges are appropriately solved.

*5.6. Prominent Challenges and Solutions*

There is a wide variety of IoT systems, from simple to complex cyber-physical systems, making it impossible to place all of the challenges and possible solutions in one table. Table 3 summarizes some challenges, important characteristics, and their possible solutions, respectively [68]. We have identified seven potential challenges and their respective BC solutions with key attributes that may be addressed before they are deployed to an IoT infrastructure [116].

**Table 3.** BC-IoT Implementation Challenges, Important Characteristics, and Possible Solutions.

| Challenges | Important Characteristics | Possible Solutions |
|---|---|---|
| Transaction Throughput | The real-time IoT data may be lost if the transaction confirmation time of public ledgers spans from 100 to 2000 TPS (Transactions Per Second). | Ripple claims to consume less time each transaction compared to BTC, ETH, Corda, and Quorum. |
| Consumption of Energy | In order to run cryptographic algorithms, IoT systems must be light-weight and have enough power. | Adaptation may be possible if manufacturing processes are planned to utilize energy. |
| Confidential Private-Key Features | To protect against eavesdropping, DTL frequently employs an asymmetric encryption strategy that takes advantage of the IoT's public key infrastructure. | Distributed IoT ledgers may be structured so that the entire ledger does not need to be replicated either. |
| Availability of the Data Transmission Space | A block size of 1 MB takes 10 min, which means that the data rate might be close to 150 MB per day. Considerable bandwidth would be required for this, and tiny IoT WANs such as Sigfox or LoRA do not have that. | Distributed IoT ledgers may be structured so that the entire ledger does not need to be replicated. |
| Congestion of the Transactions | A transaction may occur if the transaction exceeds the ledger's maximum throughput limit, which may result in increased user costs. Even with the limit provided by ripple or ETH, the real-time requirement is still not met. | The non-mining tangle-based IOTA's zero-fee transaction technique might be used. |
| The Cost of Mining and the Volatility of the Price | IoT devices that are sensitive to power consumption may not be able to use public BCs because they require high-priced hardware that relies on high-power computing. | Low-power consensus, private BCs, and non-mining DTL are all viable solutions. |
| Storage and Scalability of the Data Chain | In January 2019, BTC, ETH, and IOTA had each surpassed 200 GB, 125 GB, and 25 GB in size, indicating that the volume of data that would need to be stored to support 75 billion intelligent devices will become increasingly difficult to handle. | Distributed ledgers and big-data handling solutions might help to alleviate the problem. |

## 6. Use Case Analysis

The emerging application of distributed ledgers for BC technology can be divided into three categories: areas with common IoT controls, areas where IoT is suitable, and areas with efficient IoT solutions, according to the research on BC and distributed ledgers conducted by GSMA in collaboration with several mobile operators [117]. Figure 11 shows a comparison of different application areas, where six application areas (e.g., support compliance, device identity, data sharing, access control, micro-payments, and the supply chain) are considered for BC-based IoT use cases following assessments made by 10 operators, each with different priorities. The priority of interest of the operators were divided into three categories: minimum, medium, and maximum. For data-sharing applications, three different operators stated that it should be a minimum or medium priority, while five operators stated that it should be the highest priority, as shown in Figure 11. On the other hand, all operators stated that the access control application was a minimum priority. Furthermore, not all operators stated that micro-payment applications were a medium priority. Rather, five operators stated either a maximum or the minimum priority. For support compliance and device identity applications, five operators stated that it as a medium priority. According to GSMA, the dataset was generated by sincerely exploring all operators, but more investigation is needed before it can be used for technical and industrial purposes. Apart from these applications, Blockchain is also used for software-defined IoT infrastructures [118]. Similar work can be found in [119]. The next sections address most important four use cases that are closely related to performance, security, and scalability.

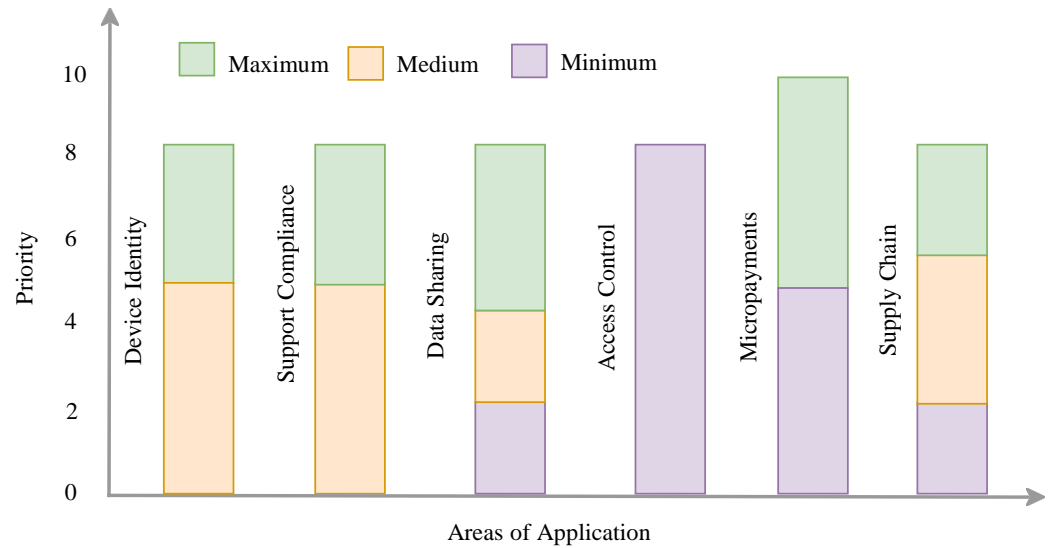

**Figure 11.** Application areas of six considered BC-based IoT use cases following assessments made by 10 operators with different priorities.

### 6.1. Use Case: Finding Appropriate IoT Devices

Device credential retrieval and tracking has been an important aspect in IoT enabling. Examples of intelligent IoT devices may be found in the following cases.

- **Case 1**: For authentication reasons, the original data and the current state of the device are stored. For example, it is important to verify the serial numbers supplied to ensure that the manufacturing firm or party is accredited by a third-party quality assurance body.
- **Case 2**: The ledger's metadata are used to verify the authenticity of software upgrades from trusted sources.
- **Case 3**: Personal data such as hardware configurations, software versions, and boot code installations should be preserved to maintain privacy.

### 6.2. Use Case: Managing IoT Access Control

In order to retain access control data for physical and virtual resources, an IoT network monitoring and recording system is unavoidable. The following are some examples of possible applications.

- **Case 1**: The ledger is used by a virtual file sharing server to protect the identity of persons and apps by encrypting access privileges for printing, saving, and editing. For example, for clients purchasing something online while away from home and cannot receive it, adopting a distributed ledger rather than a key, address, or other potentially abused code can be an advantage.
- **Case 2**: The ledger's metadata are used to verify the authenticity of software upgrades from trusted sources.

**Table 4.** Key advantages and BC applicability for the IoT finding use case.

| Use Case | Finding Appropriate IoT Devices |
|---|---|
| Key Advantages | Ensuring consistent device identification information is important for preventing vulnerabilities to unexpected third-party surveillance attacks. It can also help to keep track while adding new devices to the ledger. Any company or entity with an interest and with legitimate rights can obtain the necessary information before making a deal. |
| BC Applicability | Both public and protected BC technologies could be resilient as an application in response to the cases mentioned. Sovrin with a zero knowledge proof allows users to assert their own identity information without disclosing data directly through the ledger. |

**Table 5.** Key advantages and BC applicability for the IoT access control use case.

| Use Case | Managing IoT Access Control |
|---|---|
| Key Advantages | Limiting resource access for a specific time using a generalized API solution using smart contract rules. Access could be monitored and stored or temporarily locked by using immutable traceability to ward off illegitimate requests and to keep information for later use. Better availability and attack resilience could be achieved by copying the permission among participating nodes. |
| BC Applicability | Public BCs supporting smart contracts such as Ethereum and crypto projects such as Sovrin are able to build access management and privacy. HLF Fabric supports smart contracts such as chain-code approaches that can easily solve access control scenarios as discussed. |

### 6.3. Use Case: Supporting the Compliance of Smart Contracts

There are several situations involving various organizations in which it is crucial to determine whether or not all of them are being effectively complied with. Thus, BC smart contracts may be used to quickly and effectively enforce compliance. The following are some cases of possible applications.

**Case 1**: Distributed ledgers can be used by individuals who share personal data with their healthcare provider to ensure that only authorized medical personnel have access to that information. Ideally, the pharmacy and the general practitioner in a multiparty system should only communicate the patient's blood pressure readings in order to facilitate the dispensing of recommended medications.

**Case 2**: If a flight is delayed by 30 min, an individual may have to pay an additional $2 for airport cab service. Upon arrival, the smart contract may detect whether the additional premium has been paid in full or not in the event of micro-insurance premiums, e.g., due to a reduced cost feature of the service delivery in the smart contract. For all of the problems raised in the use cases, BC technology may be an effective solution.

**Case 3**: There must be verification of one's driving credentials, such as a valid driver's licence and a clean criminal history record, before one may drive a linked automobile. Even the automobile itself may submit trip data, service history, and even self-reported defects. One of the most efficient ways to gather data in a situation where hundreds of thousands of people are involved is to use a smart contract and BC technology.

**Table 6.** Key advantages and BC applicability for the supporting smart contract compliance use case.

| Use Case | Supporting the Compliance of Smart Contracts |
| --- | --- |
| Key Advantages | Smart contract data is immutable, so tricky mileage changes could easily be prevented with necessary transparency. For example, the journey transaction will only be added to the ledger if the odometer reading at the end is greater than the initial record. |
| BC Applicability | Public BCs such as Ethereum or open source projects such as HLF could be applied. Given that the ledger is not competing with the resource, permissioning administration, and transaction fee exemption, IBM HLF Fabric is better suited for this type of scenario. Ripple seems to be scaling in the Visa payment system. However, applying IOTA could be more useful in a micro-payment case such as Case 1, as it is designed to suit the necessarily required IoT scalability. |

### 6.4. Use Case: Maintaining Data Integrity and Confidentiality

In a distributed ledger paradigm, it is frequently hoped that data exchange while maintaining sufficient confidentiality is conceivable [120]. The ability to retain the sequence of digital signatures and data hashes provided by BC may be used to assert data integrity and IoT-related data effectively. The following are some examples of possible applications.

- *Case 1*: The manufacturing company's servers are expected to receive data from IoT devices. For instance, an intelligent thermostat linked to cloud services can provide data to the firm concerning component wear when it chooses when to turn on and off based on the current weather situation. This problem can be addressed using existing solutions, such as public key infrastructure (PKI)-driven approaches, but BC appears to be more efficient in preventing the need to reinvent procedures with regard to integrity and privacy.

- *Case 2*: An alarm system for a home or workplace may be managed by a variety of people with varying levels of access credentials. If intruders obtain access to it, law enforcement officials may need to use remote access to investigate. A distributed ledger might be particularly beneficial in this situation, which involves millions of devices being interconnected [107].

- *Case 3*: An individual may want their health care data to be shared with a researcher or medical staff through a personal fitness tracker. As a result, an individual may be ready to pay a micro premium for services provided by a manufacturer. When smart houses feature weather station/air monitoring IoT products that are shared by many parties, users similarly may consent to this kind of information sharing. A distributed ledger may be the only option for a network of machine manufacturers, practitioners, and researchers that appears to be unmanageably vast.

- *Case 4*: As a micro-generator, e.g., in a wind turbine, BC may be integrated into smart power grids to record the entire quantity of energy generated and then be used to calculate net supplier payments. Using a distributed ledger and a smart contract, it is possible to ensure that payments are made on time and in accordance with the agreed-upon rate.

**Table 7.** Key advantages and BC applicability for the data integrity and confidentiality use case.

| Use Case | Maintaining Data Integrity and Confidentiality |
|---|---|
| Key Advantages | In contrast with mobile or web applications based on relational databases, which demand operation and development efforts, distributed ledgers can easily maintain ledgers with multiple parties. There is no need for them to develop their own bespoke API either. The common API and functions of the distributed ledger save time and effort, and no extra scalability is required to ensure data integrity, security, and privacy. |
| BC Applicability | Though public BCs such as BTC and ETH show inefficiency, the directed acyclic graph-based IOTA is able to meet the challenges considering scalability required for a micro-payment system and for data sharing with integrity. Linux's open source project, namely HLF Fabric, is also able to ensure data sharing and integrity. |

## 7. Discussion

Disruptive innovations always elicit a tremendous deal of discussion and debate. Despite the fact that there are many opponents of virtual currencies, it appears unassailable that the technology that underpins them represents a major step forward in technical development. BC is a technology that is here to stay. However, there are hazards, such as updating the technology without fully insuring its operation or applying it to scenarios where the cost of the improvement does not outweigh the cost of its modification. As a result, the advantages of using BC technology with IoT applications should be thoroughly considered and approached with prudence. For the purpose of achieving a successful collaboration between BC technology and IoT applications, this study gives an overview of the major hurdles that both technologies must overcome. We have identified the critical areas in which BC technology may assist in the improvement of IoT applications. In addition, an evaluation has been presented to demonstrate the viability of using BC nodes on IoT devices. For the purpose of completing the study, existing platforms and applications were also analyzed, providing a comprehensive picture of the interplay between BC technology and the IoT paradigm. It is expected that BC technology will revolutionize IoT devices. The integration of these two technologies should be addressed, taking into account the challenges identified in this paper. The adoption of regulations is key to the inclusion of BC technology and IoT devices. This adoption could speed up the upcoming fourth industrial revolution. Consensus will also play a key role in the inclusion of the IoT as part of the mining processes and in distributing more BC technology. Nevertheless, a dualism between data confidence and the facilitation of the inclusion of embedded devices could arise. Lastly, beyond throughput, scalability, latency, and storage capacity, which affect both technologies, research efforts should also be made to ensure the security and privacy of these critical technologies.

## 8. Conclusions

The usage of BC technology is an emerging area of research, aimed at the development of efficient and scalable solutions for heterogeneous IoT applications. There is considerable concern about how efficiently BC technology can be integrated with common IoT devices while maintaining maximum throughput and privacy. In this manuscript, we have introduced different existing BC platforms and the key challenges involved in integrating them with IoT applications. In addition, this paper also provides a comprehensive analysis of how different BC platforms (e.g., BTC, ETH, and IOTA) can be used in IoT applications. Finally, we discuss some relevant use cases regarding the IoT's leading BC technology that could be helpful. It is concluded that all of the technologies discussed have great potential as a development platform aimed at enabling the efficient and real-time deployment of heterogeneous smart devices on a distributed network. In all, the IOTA technology is an open-source distributed ledger and cryptocurrency designed for IoT devices, which is more efficient in solving transaction latency and mining reward issues by saving costs and increasing performance. Furthermore, as public, private, and protected BCs each have their own set of benefits and limitations in various situations, further studies may be conducted to precisely pinpoint the pitfalls. If the challenges that arise, e.g., real-time automation and secure data processing, can be minimized, BC could be a driving force for a future driven by secured technology,

### 8.1. Theoretical Implications

Through insights on BC and its applicability to IoT, BC researchers and developers in the industry can make informed decisions about its integration into other systems. This research shows that whether a system needs BC. For a centralized solution, BC will not add any value. In addition, different consensus mechanisms are discussed to understand what sort of consensus seems applicable to a given problem. The comparison appears to provide conclusive evidence that private and consortium BCs are better suited for IoT security applications. In addition, various types of applications, such as IBM's Watson and Microsoft Azure, are discussed so that researchers can gain practical knowledge in the domain. Furthermore, apart from BC, IOTA, which is a BC-like solution but not BC in nature, is also discussed. IOTA acts as an alternative means of IoT security and seems to have higher performance because of its DAG ledger structure. Finally, industry standard use cases with specific problems are presented to provide insight into how BC integration can cause several problems. It has many advantages and can be used in many different ways, which should help researchers and developers in the field.

### 8.2. Practical Implications

In terms of the practical implications of our findings, future researchers in this field can use the findings of this study to develop new BC-based IoT applications. Furthermore, researchers should be aware of the privacy and security issues that can result from the failed integration of these technologies or their misuse. In addition, companies can use our results to better understand users' appreciation of the security of BC-based IoT connected devices, improve their products, or help users thoroughly understand the risks of the excessive use of such devices.

### 8.3. Limitations and Future Research

There are three major limitations to this study that could be addressed in future research. First, the study focuses on only six performance parameters of BC-enabled IoT applications (e.g., storage capacity, throughput, latency, privacy and security, scalability, and execution time). In the future, we will consider more performance metrics related to these heterogeneous applications. The second limitation is related to the BC technology. In this paper, we consider only ETH, BTC, and HLF. In the future, we will also consider some of the latest BC technologies, some of which are in the development phase. Finally, we consider only two existing workbenches for BC-enabled IoT applications (e.g., the MS-Azure IoT workbench and the IBM BC-integrated IoT workbench). In the future, we will look into more workbench techniques for these heterogeneous applications.

Furthermore, in further research, it is necessary to focus on improving the analysis processes used in this study as well as identify new issues related to the safety of BC-enabled IoT devices and user privacy in smart living environments.

**Author Contributions:** Conceptualization, Z.R. and X.Y.; methodology, Z.R.; software, S.T.M.; validation, Z.R. and X.Y.; and Z.R.; formal analysis, Z.R.; investigation, R.I.; resources, Z.R.; data curation, S.T.M.; writing original draft preparation, Z.R.; writing review and editing, Z.R. and A.K.; visualization, B.R.; supervision, X.Y. and A.K.; project administration, X.Y.; funding acquisition, X.Y. All authors have read and agreed to the published version of the manuscript.

**Funding:** This work was supported by the RMIT Research Stipend Scholarship (RRSS) Program. The work of Xun Yi was supported in part by the Project "Privacy-Preserving Online User Matching" under the grant ARC DP180103251. "The APC was funded by RMIT Research Stipend Scholarship (RRSS) Program".

## Abbreviations

The following abbreviations are used in this manuscript:

| | |
|------|-----------------------------------------------|
| BC | Blockchain |
| HLF | Hyperledger |
| ETH | Ethereum |
| BTC | Bitcoin |
| IoT | Internet of Things |
| MS | Microsoft |
| ECDSA | Elliptic Curve Digital Signature Algorithm |
| ZKP | Zero-Knowledge Proof |
| DP | Differential Privacy |
| CMV | Cryptographic Message Verification |
| IBM | International Business Machines Corporation |
| PK | Public Key |
| TX | Transaction |
| PoW | Proof-of-Work |
| PoS | Proof-of-Stake |
| BFT | Byzantine Fault Tolerance |
| PoA | Proof of Authority |
| DAG | Directed Acyclic Graph |
| SHA3 | Secure Hash Algorithm 3 |
| PBFT | Practical Byzantine Fault Tolerance |
| REST | Representational State Transfer |
| API | Application Programming Interface |
| MQTT | Message Queuing Telemetry Transport |
| SDK | Software Development Kit |
| YCSB | Yahoo Cloud Serving Benchmark |
| PKI | Public Key Infrastructure |
| HFC | Hyperledger Fabric Client |
| SDK | Software Development Kit |
| DLT | Distributed Ledger Technology |
| HYPF | Hyperledger Fabric |
| XFT | Cross Fault Tolerance |
| DHT | Distributed Hash Table |
| SPOF | Single Point of Failure |